\documentclass[export,twoside]{jots}
\newcommand{\victim}{
\begin{figure}[hbt!]
  \centering
  \includegraphics[width=\linewidth]{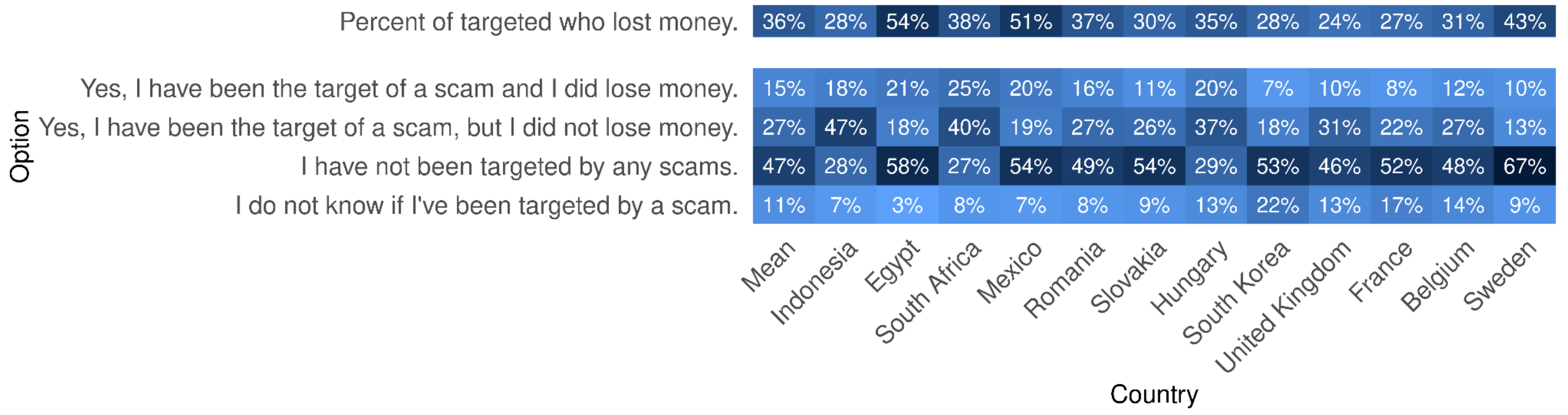}
  \caption{Weighted within-country proportions of responses to Q1: \textit{In the past year, have you been the target of a scam?} Darker colors gradually indicate higher values. The top row describes the percent who lost money among those who experienced a scam; it is computed from the two rows immediately below it.}
  \label{fig:victim}
\end{figure}
}

\newcommand{\type}{
\begin{figure}[hbt!]
  \centering
  \includegraphics[width=\linewidth]{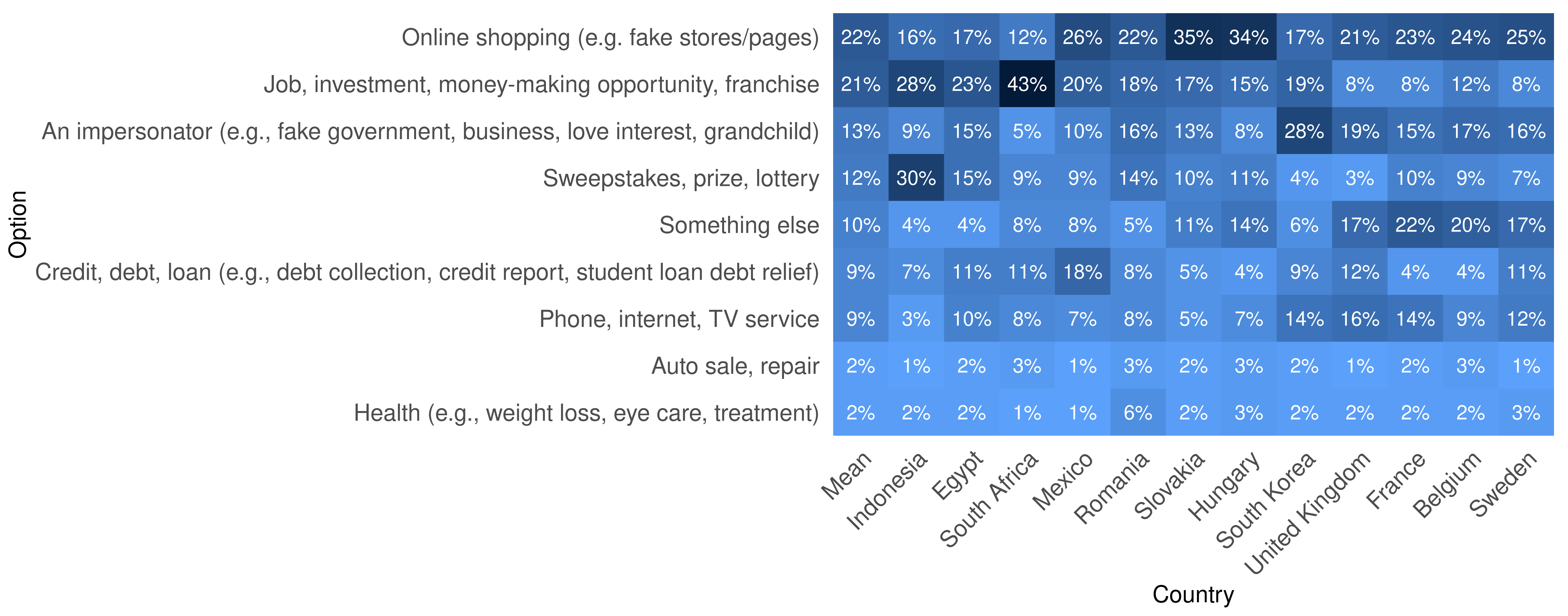}
  \caption{Weighted within-country proportions of responses to Q2: \textit{What kind of scam were you targeted by? Choose the one that most accurately describes your experience.} Darker colors gradually indicate higher values.}
  \label{fig:type}
\end{figure}
}

\newcommand{\contact}{
\begin{figure}[hbt!]
  \centering
  \includegraphics[width=\linewidth]{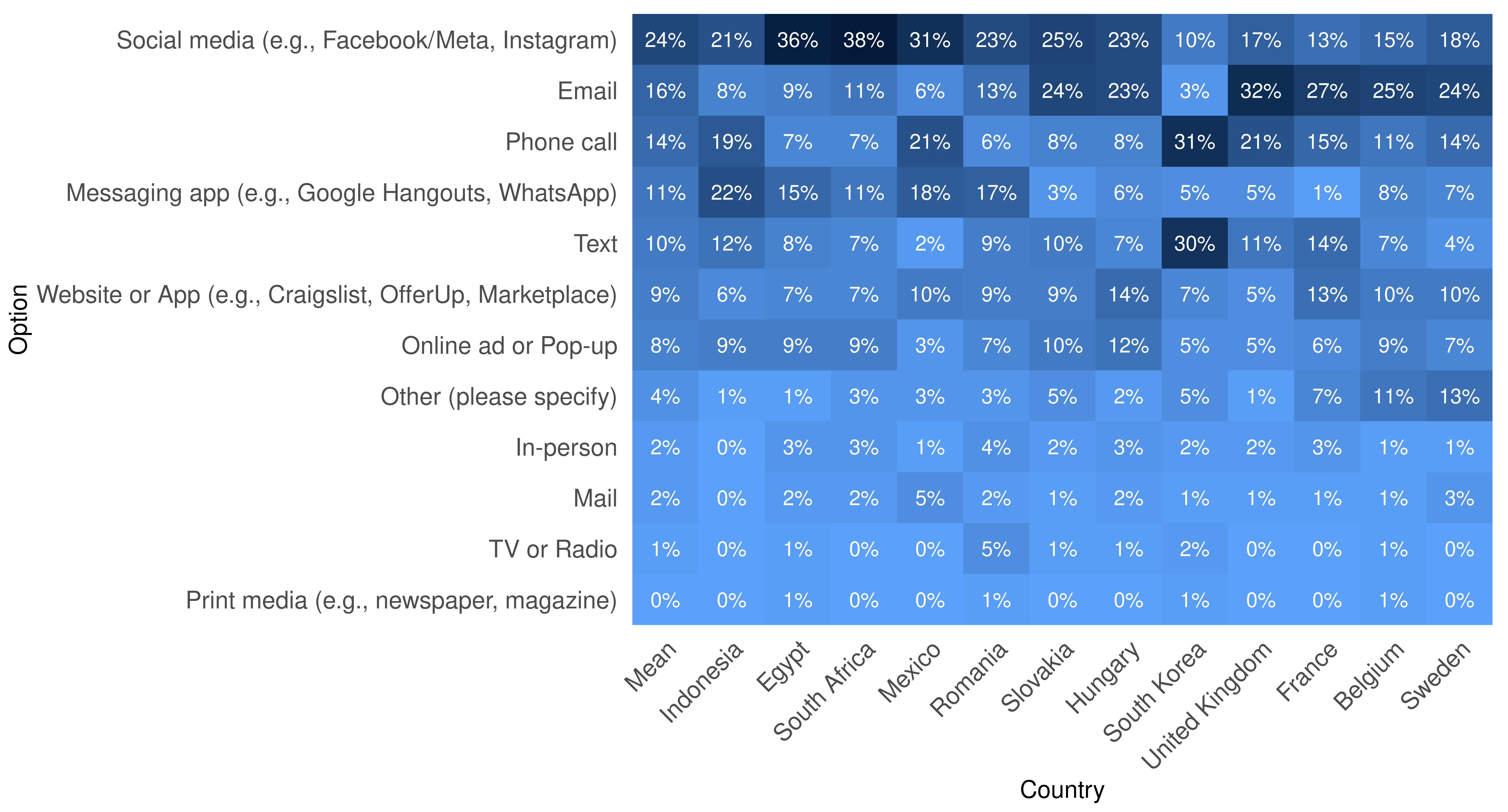}
  \caption{Weighted within-country proportions of responses to Q3: \textit{How did it start (e.g., how did they first contact you, where did you see an ad)?} Darker colors gradually indicate higher values.}
  \label{fig:contact}
\end{figure}
}

\newcommand{\payment}{
\begin{figure}[hbt!]
  \centering
  \includegraphics[width=\linewidth]{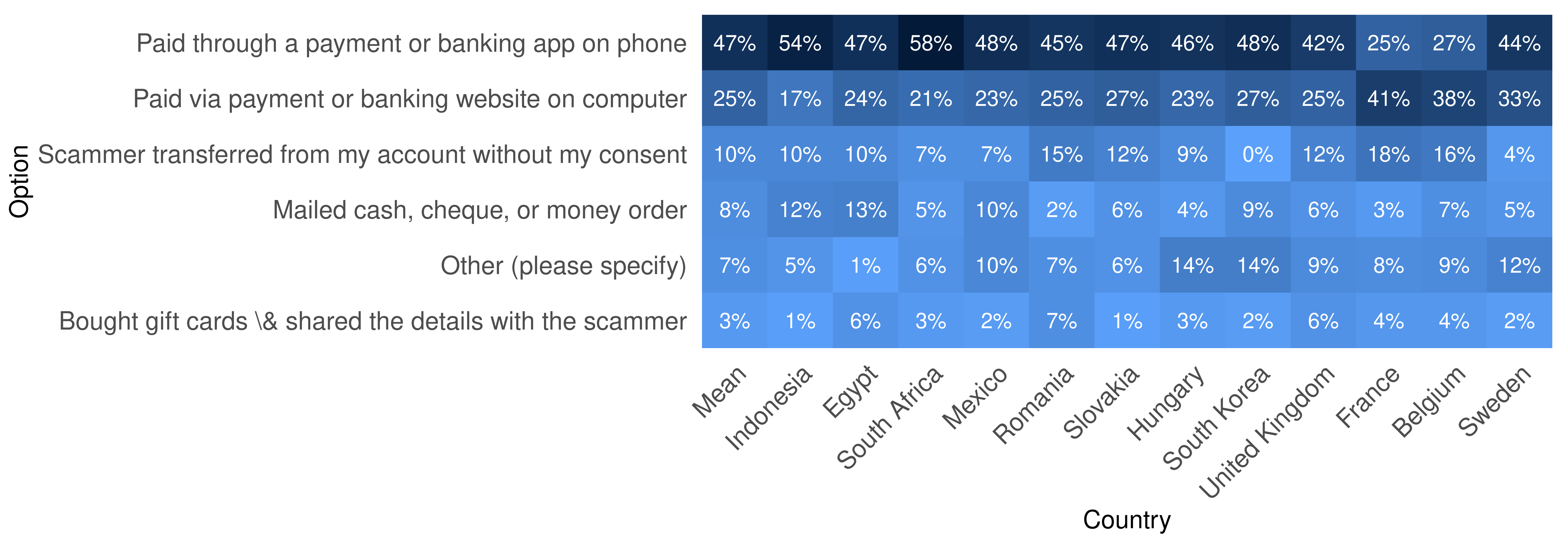}
  \caption{Weighted within-country proportions of responses to Q4: \textit{How did you pay or send the money?} Darker colors gradually indicate higher values.}
  \label{fig:payment}
\end{figure}
}

\newcommand{\reported}{
\begin{figure}[hbt!]
  \centering
  \includegraphics[width=\linewidth]{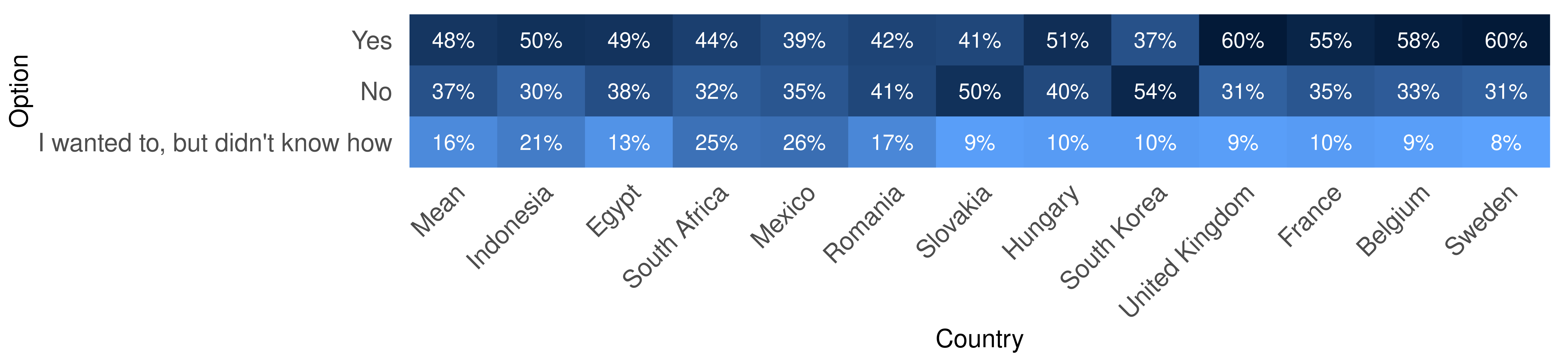}
  \caption{Weighted within-country proportions of responses to Q5: \textit{Did you report the scam to anyone?} Darker colors gradually indicate higher values.}
  \label{fig:reported}
\end{figure}
}

\newcommand{\reportwhere}{
\begin{figure}[hbt!]
  \centering
  \includegraphics[width=\linewidth]{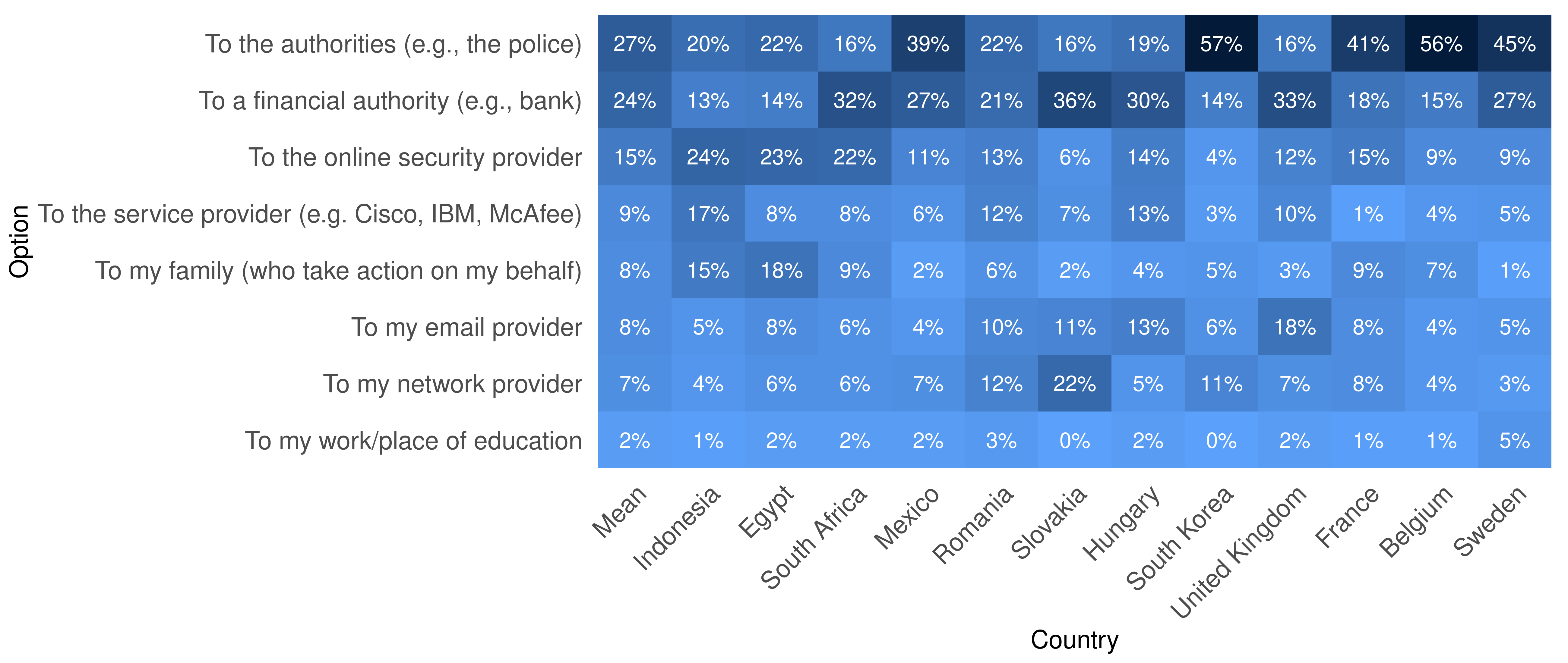}
  \caption{Weighted within-country proportions of responses to Q6: \textit{Who did you report the scam to?} Darker colors gradually indicate higher values.}
  \label{fig:reportwhere}
\end{figure}
}

\author{Mo Houtti, Abhishek Roy, Venkata Narsi Reddy Gangula, Ashley Marie Walker}
\title{A Survey of Scam Exposure, Victimization, Types, Vectors, and Reporting in 12 Countries}
\newcommand{\doi}{doi:xx.xxxx/xxxxxxx xxxx}

\begin{document}
\maketitle
\frenchspacing
\onehalfspacing

\begin{abstract}
Scams are a widespread issue with severe consequences for both victims and perpetrators, but existing data collection is fragmented, precluding global and comparative local understanding. The present study addresses this gap through a nationally representative survey (n~=~8,369) on scam exposure, victimization, types, vectors, and reporting in 12 countries: Belgium, Egypt, France, Hungary, Indonesia, Mexico, Romania, Slovakia, South Africa, South Korea, Sweden, and the United Kingdom. We analyze 6 survey questions to build a detailed quantitative picture of the scams landscape in each country, and compare across countries to identify global patterns. We find, first, that residents of less affluent countries suffer financial loss from scams more often. Second, we find that the internet plays a key role in scams across the globe, and that GNI per-capita is strongly associated with specific scam types and contact vectors. Third, we find widespread under-reporting, with residents of less affluent countries being less likely to know how to report a scam. Our findings contribute valuable insights for researchers, practitioners, and policymakers in the online fraud and scam prevention space. 
\end{abstract}

\section{Introduction}

Scams affect many people. A 2017 FTC survey found that almost 16\% of US consumers had been victims of fraud in the past year~\autocite{Anderson2019-ln}. This problem extends worldwide; a 2023 global survey by the Global Anti-Scam Alliance (GASA) found that 78\% of respondents had experienced a scam in the past year~\autocite{Abraham_undated-uy}. The material consequences of scams and fraud can be severe, extending far beyond moderate financial losses for consumers. There are many documented cases of individuals losing their life savings to scams, leading to severe material and emotional harm~\autocite{Coakley2024-cm, Khandro2024-uj, Farivar2022-jj, Kilmer2023-zn}. In some cases, businesses have been destroyed by losing immense sums of money in a scam~\autocite{Albright2023-wt}. But even these facts focus on only one set of victims. The United Nations estimates that hundreds of thousands of human trafficking victims are forced by criminal organizations to carry out scams, often after being lured to a foreign country with false promises of a job opportunity~\autocite{United_Nations_Human_Rights_Office_of_the_High_Commissioner2023-mp}. In essence, the prevalence of scams has created lucrative opportunities for organized crime, with severe consequences for both victims \textit{and} perpetrators.

The scope of this issue will likely become more important as internet access and the availability of digital payments continue to increase~\autocite{Fico_undated-ir}. The growing prevalence and widespread access to generative AI tools, such as voice cloning and high fidelity image manipulation technology, could also exacerbate the scam problem even further.  For example, in a recent incident, perpetrators used AI-generated voice and images to impersonate a company's CFO, successfully tricking a Hong Kong finance worker into wiring \$25.6 million to fraudulent bank accounts~\autocite{Ianzito2024-gv}. Proliferation of such tools could enable scams to be a lot more believable and harder to discern for users ~\autocite{Bethea2024-dn}. Unless curbed, this could increase the likelihood of victimization, and potentially reduce trust in online interactions and transactions as a whole~\autocite{FCC2024-dj}.

To help better understand and combat this problem, governments around the world regularly collect data on scams in their local jurisdictions. The Federal Trade Commission (FTC), for example, conducts surveys and publishes detailed reports~\autocite{Anderson2019-ln, noauthor_2023-dp} to map the landscape of scams in the United States. The European Commission assumes similar responsibilities in the EU~\autocite{European_Anti-Fraud_Office_undated-qa}, and analogous data collection efforts happen in places like the United Kingdom~\autocite{Jones2022-no}, Australia~\autocite{abs_personal_2024}, and South Korea~\autocite{Jun-hee2023-jo}.

But despite these efforts, we do not have a good quantitative understanding of scams at a global level. Because data collection typically happens \textit{within} jurisdictions, methodological differences make it impossible to compare findings across borders. While some global reports do exist, they either compile data from existing sources or use survey methodologies that do not achieve population-representative samples (e.g.,~\textcite{Abraham_undated-uy, Lewis2020-kt}). Some industry firms (e.g.,~\textcite{NortonLifeLock2022-wi}) have conducted representative global surveys, but these have tended to be more high-level and are therefore insufficiently granular to let us draw precise and practical conclusions about the global scams landscape. This in turn hinders policymakers, practitioners, and researchers from identifying and targeting scams in contextually appropriate ways.

To address this gap, we conducted a nationally representative survey (n = 8,369) in 12 countries---Belgium, Egypt, France, Hungary, Indonesia, Mexico, Romania, Slovakia, South Africa, South Korea, Sweden, and the United Kingdom. The survey asked people about their experiences with scams, including the type of scam they had most recently experienced, the vector through which they experienced the scam, whether they had sent the scammer money, and whether/where they reported the scam. Importantly, we ask these questions through a global and nationally representative survey, making it possible to draw true comparisons and contrasts between countries. This provides a couple key advantages. First, it lets us contextualize local findings. If we find that 25\% of South Africans report having lost money from a scam in the past year, is that a lot? According to our survey, the answer in this case is yes---but this argument would be much more difficult to make in the absence of global data. Second, while not a perfect substitute for \textit{comprehensive} global data, analyzing scam data across a broad spectrum of countries can help resource-constrained governments prioritize their efforts. If consistent scam types emerge across diverse economies, or if similar economies suffer from specific scam types, they can be valuable signals to guide local data collection and enforcement efforts. Concretely, the survey let us answer the following research questions in multiple countries:

\begin{itemize}[noitemsep]
        \item RQ1: How common are scam exposure and victimization?
        \item RQ2: How prevalent are specific scam types?
        \item RQ3: To what extent are scams technology-mediated?
        \item RQ4: How comprehensive are reporting-based data sources about scam victimization?
\end{itemize}

We present several useful findings, from which we contribute concrete takeaways for researchers, practitioners, and policymakers in the online safety and fraud prevention spaces. First, we find that residents of less affluent countries suffer financial loss from scams more often. Second, we find that the internet plays a key role in scams across all surveyed countries, but that GNI per-capita is strongly correlated with specific scam types and contact methods from scammers. For example, money-making scams are more common in less affluent countries. Third, we find widespread under-reporting of scams. On overage, half of those exposed to a scam in a given country do not report it at all---and residents of less affluent countries are more likely to not know how to report a scam.

\section{Background}

\subsection{Scam Exposure and Victimization}

Scams are a growing issue, with losses from reported scams increasing at a rapid rate. Industry surveys and reports compiled by law enforcement are the primary sources of scam prevalence data. The FBI's Internet Crime Report 2023 showed that losses grew at upwards of 20\% year over year in the last three years~\autocite{fbi_ic3_2023}. The Global Anti-Scam Alliance (GASA)'s State of Scams 2023 report estimated that losses from scams topped USD 1 trillion in 2023 and accounted for about 1\% of global GDP~\autocite{Abraham_undated-uy}. While estimates for exposure to scams vary widely across surveys and countries, it is generally accepted that the exposure to scams is increasing year over year. GASA's 2023 report stated that ~78\% of participants experienced at least one scam in the past 12 months. 

The GASA report estimated that developing countries like Kenya, Vietnam, Thailand and Brazil lost more than 3\% of their GDP to scams~\autocite{Abraham_undated-uy}. A 2020 survey commissioned by the European Commission (EC) found that the exposure to fraud is generally higher in ``connected countries'' identified as countries with a relatively high proportion of individuals making internet purchases~\autocite{ec_scams_survey}. In the context of online scams, increased internet activity (e.g., online purchases) expands the pool of potential targets and the opportunities for motivated offenders~\autocite{kigerl2012routine, felson1980RAT, miro2014routine}. Prior research indicates that a variety of psychological (e.g. overconfidence, optimism bias), socio-demographic (e.g. age, education) and situational factors (e.g. emotional distress, financial strain) play a role in susceptibility to scams~\autocite{Whitty2019-hv,Modic2013-wn,Button2014-ip}.

In this study, we looked at how the exposure and victimization to scams vary across geographies.

\subsection{Scam Types and Technology’s Role in Scams}

Scams are heterogeneous and evolve based on a variety of factors such as emerging technologies (e.g., cryptocurrency), changing communication channels (e.g., social media), and global events (e.g., the COVID-19 pandemic, the Russia-Ukraine war)~\autocite{xu2022differentiating}. Scammers adapt their methods based on the individual circumstances of their targets, ranging from preying on financial vulnerabilities through lottery scams to exploiting loneliness through romance scams. While attempts have been made by multiple government agencies and researchers, there is still no universally accepted classification of scams. DeLiema et al. classify consumer fraud into four categories: opportunity-based scams, threat-based scams, consumer purchase scams, and phishing scams~\autocite{DeLiema2023-yo}. In contrast, FINRA's Financial Fraud Research Center has developed a framework for a taxonomy of fraud modeled after international crime classification systems~\autocite{beals2015framework}. In this study, we used a scam classification based on the role scammers play and the service they purportedly offer to understand the prevalence of such scams. 

As stated earlier, scammers take advantage of technological advances to effectively reach and scam their targets. Scammers use a variety of communication channels such as phone calls, text messages, emails, social media, and mobile apps to establish contact with their targets. Scammers also leverage advances in payment methods with the emergence of irreversible payment methods such as real-time payments, cryptocurrency, digital wallets, etc. Notably, scams utilizing cryptocurrency have skyrocketed, with the FTC reporting that cryptocurrency scams make up more than 85\% of losses due to investment scams~\autocite{noauthor_2023-dp}. In this study, we looked at how scammers use technology to reach their targets and receive payments across countries of interest.

\subsection{Limitations of Reporting-based Data Sources}

Despite scams leading to devastating financial losses and causing enormous psychological distress to victims~\autocite{munton2023scams}, they are highly under-reported to authorities with the FTC estimating that only 10-12\% of scams are actually reported to them~\autocite{noauthor_2023-dp}. ~\textcite{Deliema2019-bw} show that scams are under-acknowledged and under-reported even in survey research, likely due to social desirability bias or refusal to acknowledge victimization. In the limited studies that surveyed known victim pools, only about half of known fraud victims admitted to being defrauded~\autocite{button2009fraud,finra_sr_fraud_risk_survey}. This shows that survey-based estimates are better than reporting-based estimates to assess scam prevalence and victimization.

While reports like those from the FBI and industry players provide valuable information, they suffer from methodological disadvantages that hinder their representativeness. Reports based on complaint data do not provide a complete picture, as scams are severely under-reported to government agencies. Some surveys by industry players are global but often focus on a narrow aspect of the problem such as real-time payments based scams or text message scams, or have representativeness issues. Through this survey, we looked at scam exposure, victimization and reporting attitudes across regions using a representative sample of participants in order to identify regional disparities and inform targeted prevention and intervention strategies. 

\section{Methods}

\subsection{Survey Overview}

The questions analyzed in this paper were part of a larger survey covering a variety of digital safety issues. The survey was deployed through Morning Consult (MC)---a leading survey research firm---using a mixed-panels approach, which leverages the strengths of different data collection methods to create a more comprehensive and representative picture of public opinion. MC uses roughly 55 survey panel providers to conduct interviews across numerous countries. This panel network provides access to tens of millions of survey respondents via recruitment from thousands of websites, mobile apps, social networks, email lists, and publishers. They use a wide mix of panel providers with different recruitment methods to diversify their respondent pools and ensure maximal access to different respondent groups. Recruited participants completed the survey online, using mobile or desktop devices.

12 countries were surveyed: Belgium, Egypt, France, Hungary, Indonesia, Mexico, Romania, Slovakia, South Africa, South Korea, Sweden, and the United Kingdom. Questions were translated by teams with native speaker(s) in the target languages through a process consisting of translation, editing, proofreading, and quality assurance. Demographic factors used for stratified sampling and weighting varied by country to account for contextual factors such as the availability of reliable census data. Weights and sample targets were based on general population proportions for adults (18 and older) in Belgium, Egypt, France, Hungary, Indonesia, Romania, Slovakia, South Korea, Sweden, and the UK. Mexico and South Africa have internet access penetration below 80\%, and folks without access to the internet are more likely to fall into lower education demographics. Therefore, to avoid over-representing adults with lower education, weights and sample targets for those two countries were based on internet population proportions for adults (18 and older). Responses were weighted using standard raking procedures~\autocite{Battaglia2009-ad}. Table~\ref{tab:surveytab} reports the demographic factors used to derive sample targets and weights in each country.

\begin{table}[hbt!]

\caption{Demographic factors used for stratified sampling and weighting in each country.}
\label{tab:surveytab}
\small
\begin{tabular}{llll}
\toprule
Country                   & Sample size & Sample targets based on & Weighting dimensions           \\
\midrule
Belgium                   & 504 & Age, gender             & Age, gender, education, region \\
Egypt                     & 740 & Age, gender             & Age, gender, education, region \\
France                    & 738 & Age, gender             & Age, gender, education, region \\
Hungary                   & 527 & Age, gender             & Age, gender                    \\
Indonesia                 & 752 & Age, gender             & Age, gender, education, region \\
Mexico                    & 778 & Age, gender, education  & Age, gender, education, region \\
Romania                   & 761 & Age, gender             & Age, gender, education, region \\
Slovakia                  & 724 & Age, gender             & Age, gender                    \\
South Africa              & 773 & Age, gender, education  & Age, gender                    \\
South Korea               & 782 & Age, gender             & Age, gender                    \\
Sweden                    & 533 & Age, gender             & Age, gender, education, region \\
United Kingdom            & 757 & Age, gender, education  & Age, gender, education, region \\
\bottomrule
\end{tabular}
\end{table}

\subsection{Survey Questions}

We used the FTC's online fraud reporting tool~\autocite{Federal_Trade_Commission_undated-hm} as a guide when formulating our survey questions. This let us ensure that the questions captured the kind of information about scams that is useful to government institutions and that the multiple-choice options adequately covered common scam experiences. To account for the likelihood of multiple scam experiences, respondents were instructed to answer questions based on their most recent scam experience (if any) in the past year. We reasoned that respondents might have trouble recalling details of earlier scam experiences, making their reports less reliable. Obtaining a cross-section of most recent scam experiences would also closely approximate the breakdown of scam experiences overall while letting us simplify the survey for respondents.

While ``scams'' and ``fraud'' technically have different definitions, the two terms are often used interchangeably to refer to attempts to trick someone for monetary gain. Indeed, the aforementioned FTC reporting tool~\autocite{Federal_Trade_Commission_undated-hm} does not distinguish between the two. Rather than providing a narrow definition that might unintentionally exclude instances commonly considered to be scams, we decided to mirror the FTC and rely on respondents' colloquial understanding of the term.

Recall that we articulated four research questions. We state each research question before its corresponding survey questions.

\textbf{RQ1:}~\textit{How common are scam exposure and victimization?}

We analyzed responses to Q1 to understand the frequency with which internet users are exposed to and materially harmed by scams:

\textbf{Q1: \textit{In the past year, have you been the target of a scam?}}
\begin{itemize}[noitemsep]
    \item Yes, I have been the target of a scam, but I did not lose money.
    \item Yes, I have been the target of a scam and I did lose money.
    \item I have not been targeted by any scams.
    \item I do not know if I've been targeted by a scam.
\end{itemize}

\textbf{RQ2:}~\textit{How prevalent are specific scam types?}

We analyzed Q2 to understand the types of scams internet users experience:

\textbf{Q2: \textit{What kind of scam were you targeted by? Choose the one that most accurately describes your experience.}}
\begin{itemize}[noitemsep]
    \item An impersonator (e.g., fake government, business, love interest, grandchild)
    \item Job, investment, money-making opportunity, franchise
    \item Phone, internet, TV service
    \item Health (e.g., weight loss, eye care, treatment)
    \item Online shopping (e.g. fake stores/pages)
    \item Sweepstakes, prize, lottery
    \item Auto sale, repair
    \item Credit, debt, loan (e.g., debt collection, credit report, student loan debt relief)
    \item Something else
\end{itemize}

\textbf{RQ3:}~\textit{To what extent are scams technology-mediated?}

Responses to Q3 and Q4 let us understand to what extent users' scam experiences are technology-mediated. Q4 was only given to respondents who indicated having directly sent the scammer money in a separate question. 

\textbf{Q3: \textit{How did it start (e.g., how did they first contact you, where did you see an ad)?}}
\begin{itemize}[noitemsep]
    \item Phone call
    \item Social media (e.g., Facebook/Meta, Instagram)
    \item Online ad or Pop-up
    \item Website or App (e.g., Craigslist, OfferUp, Marketplace)
    \item Messaging app (e.g., Google Hangouts, WhatsApp)
    \item Email
    \item Text
    \item Mail
    \item In-person
    \item TV or Radio
    \item Print media (e.g., newspaper, magazine)
    \item Other (please specify)
\end{itemize}

\textbf{Q4: \textit{How did you pay or send the money?}}
\begin{itemize}[noitemsep]
    \item Scammer transferred from my account without my consent
    \item Paid through a payment or banking app on phone
    \item Paid via payment or banking website on computer
    \item Mailed cash, cheque, or money order
    \item Bought gift cards \& shared the details with the scammer
    \item Other (please specify)
\end{itemize}

\textbf{RQ4:}~\textit{How comprehensive are reporting-based data sources about scam victimization?}

Finally Q5 and Q6 let us contextualize the current research information landscape on scams, which often relies heavily on reporting-based data.

\textbf{Q5: \textit{Did you report the scam to anyone?}}
\begin{itemize}[noitemsep]
    \item Yes
    \item No
    \item I wanted to, but didn't know how
\end{itemize}

Those who did report the scam were asked:

\textbf{Q6: \textit{Who did you report the scam to?}}
\begin{itemize}[noitemsep]
    \item To a financial authority (e.g., bank)
    \item To the service provider (e.g. Cisco, IBM, McAfee)
    \item To the authorities (e.g., the police)
    \item To my email provider
    \item To the online security provider
    \item To my network provider
    \item To my family (who take action on my behalf)
    \item To my work/place of education
\end{itemize}

\subsection{Analysis}

For each survey question, we report the weighted percentage of respondents who selected each multiple-choice option in a heatmap. Countries are ordered along the x-axis by ascending GNI per-capita in 2021 (the most recent year for which complete data was available). Where the heatmaps suggest possible GNI-based patterns among the most frequently selected or most pertinent options, we compute Spearman's rank-order correlations to verify and quantify the associations.

\section{Results}

\textbf{RQ1:}~\textit{How common are scam exposure and victimization?}

\subsection{Users in Less Affluent Countries Are at Greater Risk}

On average, 15\% of a country's internet users lost money from scams in the past year (Figure~\ref{fig:victim}). Our results also reveal a troubling pattern: internet users in less affluent countries are at greater risk of falling victim to scams. South Africa, for example, had a GNI per-capita of \$12,948 and the highest scam victimization rate at 25\%. South Korea, with a GNI per-capita of \$44,501, had less than a third of South Africa's victimization rate (7\%). We found a strong inverse correlation between GNI per-capita and scam victimization rate ($\rho=-0.73$, $p=0.007$). This is especially concerning because lower incomes in countries with lower GNI per-capita make it more difficult for victims to recover from financial loss.

\subsection{Scam Exposure Does Not Reduce Victimization Rate}

Some countries with low rates of scam exposure nevertheless have a high rate at which those who do experience scams actually suffer losses (e.g., Egypt, Mexico, and Sweden). Users in these countries may simultaneously be well protected from \textit{exposure} to scams, but not resilient against scams themselves. Indeed, one could reasonably hypothesize that less frequent scam exposure might make users less alert and therefore less resilient overall.

However, we found no correlation in either direction between the scam loss rate---the top row in Figure~\ref{fig:victim}---and scam exposure rate---the sum of the following two rows ($\rho=0.04$, $p=0.90$). This suggests a more complicated relationship between scam exposure and avoidance of financial loss from scams, with additional factors possibly at play. For example, countries where users are frequently exposed to scams may become more resilient in \textit{absolute} terms, but be targeted using more sophisticated techniques in response. This kind of tailoring would be consistent with the results we cover next, demonstrating some differences in scam types and communication vectors based on countries' affluence.

\victim

\textbf{RQ2:}~\textit{How prevalent are specific scam types?}

\subsection{Top Scam Types Are Fairly Consistent Across Countries}

The two most common scam types were online shopping and money-making scams (Figure~\ref{fig:type}). There were only two countries where neither of these was the most-selected option: Indonesia with sweepstakes/lottery scams (30\%), and South Korea with impersonation scams (28\%). In both cases, however, money-making and online shopping scams were still in second and third place, respectively.

Despite this consistency, we did see differences based on countries' economies. Money-making scams were more common in less affluent countries; GNI per-capita and the prevalence of money-making scams had a very strong inverse correlation ($\rho=-0.90$, $p<0.001$). The prevalence of online shopping scams was not significantly correlated with GNI per-capita ($\rho=0.44$, $p=0.15$). 

\type

\textbf{RQ3:}~\textit{To what extent are scams technology-mediated?}

\subsection{From First Contact to Payment, the Internet Plays a Key Role}

Consistent with FTC reports in the US~\autocite{noauthor_2023-dp}, the most common method of first contact across countries (Figure~\ref{fig:contact}) was social media (24\%), followed by email (16\%) and phone (14\%). On average, 68\% of those exposed to a scam in a country had first contact with a scammer through an internet-based technology (social media, email, messaging app, website or app, or online ad or pop up). Social media and messaging apps were more common as initial contact methods in less affluent countries, while email-based scams were more prominent in more affluent countries. GNI per-capita had a strong inverse correlation with the prevalence of social media ($\rho=-0.71$, $p=0.009$) and messaging apps ($\rho=-0.65$, $p=0.02$) as first contact methods, and a strong positive correlation with email ($\rho=0.67$, $p=0.02$). Interestingly, South Korea was a major outlier in our data; phone calls and texts were the most common contact methods there, with internet-based methods making up only 30\%.

\contact

Mobile apps were a key vector for payments to scammers. Mobile-based payments were almost twice as common as computer-based ones (Figure~\ref{fig:payment}). This pattern held across most countries---with France and Belgium being the only exceptions. This may simply reflect online banking habits across countries, but nonetheless provides good evidence that scam prevention should include a focus on mobile payments, and further substantiates reports calling out the risks of the increasing global adoption of real-time payments (RTP)~\autocite{Fico_undated-ir}. The prevalence of mobile-based payments to a scammer had a strong inverse correlation with GNI per-capita ($\rho=-0.79$, $p=0.002$), suggesting the adverse effects of RTP adoption may be more prominent in less affluent countries.

\payment


\textbf{RQ4:}~\textit{How comprehensive are reporting-based data sources about scam victimization?}

\subsection{Many Scams Go Unreported}

Our findings substantiate the widespread understanding that scams are under-reported, and provide quantitative evidence of the extent to which this is true globally. On average, over a third (37\%) of a country's scam victims did not report the scam to anyone (Figure~\ref{fig:reported}). In Slovakia and South Korea, that number was closer to half (50\% and 54\%, respectively). Users in less affluent countries more often wanted to report scams but did not know how ($\rho=-0.82$, $p=0.001$). This suggests that inferences based on victim reports are likely to suffer more under-reporting bias in resource-constrained countries, where survey data is also less likely to exist.

\reported

Reporting to government authorities constituted an even smaller fraction of these totals (Figure~\ref{fig:reportwhere}). Of those who reported scams, less than a third (27\%) reported them to the authorities. This varied widely by country---e.g., 57\% of South Koreans who reported scams reported them to authorities versus only 16\% of South Africans. GNI per-capita did not correlate significantly with reporting to either authorities ($\rho=0.47$, $p=0.12$) or financial institutions ($\rho=0.19$, $p=0.56$), though the proportions may reflect the degree to which victims in each country believe a particular institution may be able to take action for them (e.g., retrieve their lost money).

\reportwhere

\section{Discussion}

Scams are a widespread and growing problem affecting people worldwide. The recent cost-of-living crisis \autocite{Economist_Intelligence_Unit2023-nl} has exacerbated the financial impact of scams, leaving victims even more vulnerable. Greater rates of financial loss from scam victimization coupled with more difficulty recovering from financial loss (due to low income) in less affluent economies makes this issue worth even more attention. Consequently, governments, central banks, industry and civil society around the world have responded with various tools in their arsenal, from public awareness campaigns to regulations\autocite{Stainsby2023-ue}.

This lack of standardized data hinders our ability to understand and address scam victimization effectively. Without a global survey on scam prevalence and victimization, it is hard to accurately assess and describe the true size of the issue, develop effective solutions or bring appropriate resources to bear. Our work tries to fill in some of the gaps in terms of understanding who is most at risk, and to what types of scams, so regulators and industry can better target their interventions. Furthermore, as our numbers on contact method and payment method show, the scams landscapes in low GNI per-capita countries often look different, so minority-world-focused solutions are likely not going to be as effective in the majority world. This survey aims to fill that gap in literature.

In this section, we highlight some of the key drivers of scam victimization across countries as demonstrated by our findings, and define a multi-level structural approach to fight back against this issue.

\subsection{Drivers of Scam Victimization Across Countries}

While the specific tactics and methods used by scammers vary across countries, there are several common tailwinds and themes contributing to the widespread vulnerability and victimization to scams.

\subsubsection{Widespread and fast adoption of digital payment systems globally post-COVID appears to be a key driver in increased scam prevalence.}

As we show in Figure~\ref{fig:contact}, digital payment solutions are involved in a  large fraction of the scams incidents across all the countries we surveyed. This observation mirrors the growth of the total number of non-cash transactions happening globally, which is estimated to accelerate to \$2.3 trillion by 2027, growing at a rate of 15\% annually~\autocite{Capgemini2023-rc}.

Frauds and scams are not, however, unique to digital payment systems. While digital payment systems are fast, convenient and thereby have a large volume, they often offer limited recourse for victims as compared to modes of payment like credit cards. While with the adoption of such payment systems we have seen a massive growth in economic activity, this has also increased the total number of scam activities over time. For example, UK's Payment Systems Regulator (PSR) found just 0.1\% of fast payments in 2021 were fraudulent---well above the global average of 0.03\% for card transactions~\autocite{World_Bank2023-yq}. While this trend of growth in adoption of digital payments has implications for the total volume of scams occurring globally, it is particularly pertinent in Asia Pacific where non-cash transactions seem to be growing at an estimated rate of 19.8\% annually~\autocite{Capgemini2023-rc}.
The type of scams that are prevalent vary country to country, but it appears that focusing research and enforcement efforts on scams facilitated through digital payment services is a productive intervention space, especially for those countries.

\subsubsection{Scams follow where users go online, and locale specific factors are highly salient in understanding specific scam victimization.}

As we show in Figure~\ref{fig:type}, the types of scams that are popular differ by country, but there seems be a trend of scammers using hooks that are most likely to yield results in their target locales. For example, high rates of gambling related scams in Indonesia is correlated with a rapid rise in the illegal gambling industry in the country, with one PPTAK (Indonesian Financial Transaction Reports and Analysis Center) spokesperson noting that the turnover in this industry grew from 57 trillion Rp in 2021 to 81 trillion Rp in 2022~\autocite{Bhwana2023-rp}. Similarly, we note that the high rate of jobs scams in South Africa is correlated with South Africa having one of the highest unemployment rates in the world at 32.1\%, signifying that scammers are preying on a large market of job-seekers in the market~\autocite{Desiree_Manamela2024-nz}.

\subsubsection{Scams go severely under-reported, creating ecosystem-wide blind spots.}
Since globally only about half the scams are reported to authorities, the current national level estimates of scam victimization and prevalence are most likely highly underestimated (see Figure~\ref{fig:reported}). This trend of under-reporting is especially pronounced in countries like Indonesia, South Africa and Mexico, where more than 1 in 5 users reported being unable to report scams to authorities due to unclear reporting mechanisms. As we discuss further in next section, generating public awareness of reporting tools and mechanisms available to the general public could be a productive area of intervention.

The low effective rates of enforcement against scams and low likelihood of loss recovery may also play a role in discouraging victims from reporting scams to authorities~\autocite{Dolgin2023-bh}. For example, the 2023 State of Scams in Asia report by Global Anti Scam Alliance, a cross-stakeholder body, found that there was a positive correlation between the recovery of financial losses and people's willingness to report scams to authorities~\autocite{Global-Anti-Scam-Alliance2023-fb}. Another recent report by Third Way, a public policy think tank, estimated that in the US the enforcement rate for reported incidents to the Internet Crime Complaint Center (IC3) was 0.3\% \autocite{Eoyang2018-wk}.  Governments and central banks must ensure allocation of more resources to law enforcement agencies, a streamlined reporting mechanism, and ensure that people have access to proper redressal mechanisms .

\subsection{Addressing the Scam Problem: Potential Interventions}

Scam victimization erodes consumer trust, jeopardizes businesses and harms the broader economy. A multi- faceted targeted intervention approach is necessary to develop and assess the effectiveness of various intervention approaches. As we show in \textit{Section 4.3}, internet based communication methods play a critical mediating role in victimization. Prior work in this area has classified scams along the following broad categories (\cite{DeLiema2023-yo}): 
\begin{enumerate}
\item Opportunity-based scams: typically involves some monetary / social / emotional reward based hook.
\item Threat-based scams: typically involving threats of negative consequences and blackmail
\item Consumer purchase scam: typically happens during online shopping or marketplace interactions.
\item Phishing scams: often involves techniques like impersonation of a person/entity that victims trust. This is then parlayed to elicit either money or sensitive information from the victim.
\end{enumerate}
As we show in \textit{Section 4.2}, the top scam types remain fairly consistent across surveyed countries; the most popular scam channels were shopping (consumer purchase scam), job/investment (opportunity-based scam) and impersonation-based scams (phishing, romance or friends/family impersonation scams). This bodes well for an approach where we can target and educate users about the underlying manipulation techniques used in these scams, strengthen the resilience of such channels themselves, and raise overall awareness. For countries with outlier patterns, like Indonesia (high victimization from lottery scams), South Africa (high victimization from job-based scams), and South Korea (impersonation and romance scams), it would make sense to take a more white glove approach where Public-Private sector collaboration is called for to target and curb these trends.

\paragraph{Interventions to strengthen user resilience}

Recognizing that users are often the weakest link in the scam ecosystem~\autocite{Baral2019-ab}, we must prioritize strategies to enhance their resilience. A multi- faceted approach, akin to a ``swarm the problem'' strategy, is necessary to develop and assess the effectiveness of various intervention approaches. This area is ripe with opportunities to borrow from intervention tool kits developed and tested against other forms of online harms like misinformation~\autocite{Kozyreva2022-to}, for example:

\begin{enumerate}

\item \textbf{General Awareness Campaigns:} Public awareness campaigns help raise awareness of common scam tactics and warning signs, and can theoretically help the broader public to recognize and avoid potential threats~\autocite{Miller1983-md}. Having some pre-existing familiarity with scam type has been shown to help individuals better manage risks, but while general awareness campaigns are one of the most prominent intervention strategy deployed by stakeholders, its efficacy in bolstering user resiliency is mixed and unproven~\autocite{Downs2006-wz}~\autocite{Jensen2024-qh}. There is some research showing efficacy of such campaigns in raising awareness about reporting channels, and in that capacity this could be a tool of interest for stakeholders to potentially alleviate the problem of under-reporting we noticed across most countries~\autocite{Jeremy-Burke-Francisco-Perez-Arce-Christine-Kieffer-Robert-Mascio-Gary-Mottola-Olivia-Valdes2021-xv}.

\item \textbf{Behavioral Nudges and Credibility Cues:} Embedding lightweight behavioral nudges might be a good low-friction way of guiding users towards safer online practices~\autocite{Pennycook2021-gm}. For instance, research has shown that using attention prompts to verify if someone is receiving or sending money on a fast payment app can significantly reduce the risk of falling victim to scams where scammers rely on the user not paying attention and sending money instead of receiving it~\autocite{Jha2022-dp}. Furthermore, since impersonation based scams are prevalent across all surveyed countries and users rely on visual trust indicators to assess credibility online~\autocite{Jakobsson2007-wx}, another intervention worth exploring more broadly here would be providing users with additional credibility cues (e.g. verification check marks) for verified accounts across platforms. As we show in \textit{Figure \ref{fig:contact}}, since most scams begin from platforms like social media, email, phone call and text messages, adding positive credibility cues (like verification check marks) or negative credibility cues (like potential spam label) could forewarn individuals and help guide them towards making safer interaction decisions online.

\item \textbf{Inoculation-Styled Interventions:} Scammers often rely on a handful of manipulation techniques to trick their victims into falling for scams. Like vaccines, these interventions expose users to weaker forms of these techniques in a controlled environment to help them build immunity or ``mental antibodies'' to these real-world threats~\autocite{McGuire1961-yv}. Prior research has attempted to map out user journeys of victims and understand what manipulation techniques are at play, particularly for romance scams~\autocite{Whitty2013-gb}. Building on this, future research could replicate such mapping to the 3 most popular scam journeys across countries: consumer purchases online, opportunity-based scams (job/investment) and impersonation-based scams (phishing, romance and family/friend impersonation). Users can then be \textit{inoculated} against these manipulation techniques and tactics. For example, Robb \& Wendel 2023 showed that there is strong evidence that inoculation techniques increased users' ability to detect SSN scam emails (government impersonation) as compared to general tips about scams~autocite{Robb2023-gz}.

\item \textbf{Digital and Financial Literacy-based Interventions:} Since scammers prey on lack of digital \& financial literacy among new and habitual internet users, deploying low-cost comprehensive intervention programs that equip individuals with the knowledge and skills to navigate the digital landscape safely, critically evaluate information, and make informed decisions would go a long way as a preventative measure in this space~\autocite{McGrew2023-uh}. Research has shown that educational interventions can reduce susceptibility to financial fraud pitches~\autocite{Burke2022-wv}. While digital literacy is known to enhance resilience~\autocite{Graham2017-he}, the field lacks interventions comparable to those addressing misinformation~\autocite{Kozyreva2022-to}. Such tools could be powerful interventions against opportunity-based scams, particularly those that involve investment schemes preying on individuals with low financial literacy.

\end{enumerate}

\subsection{Scope for Future Research}

\paragraph{Deepening research to understand scam victimization and barriers to reporting}

To effectively combat scams, we need a more comprehensive understanding of the phenomenon. While quantitative data, such as that presented in this study, helps to gauge the scale of victimization, we also need equally powerful qualitative research to better understand the issues uncovered in this study, particularly in less affluent countries. In-depth qualitative research, particularly in the following areas, could help tailor effective interventions and policies:

\begin{enumerate}
    \item \textbf{Understanding Lived Experiences of Scam Victims:} Scam victimization leaves behind profound and lasting scars. Beyond immediate impact like financial hardships, and emotional repercussions like feelings of shame, anger, and betrayal, it can lead to long-term instabilities and victims socially isolating themselves resulting from a loss of trust in those around them~\autocite{Cross2018-ht, Button2014-bj, Whitty2016-gm}. But while qualitative research related to lived experiences of victims in more affluent countries has been insightful, experiences of victims in non-affluent countries largely remains a blind spot. Findings from \textit{Section 4.1} in this study underscore an urgent need for such studies as GNI per-capita and scam victimization rates were strongly negatively correlated. Understanding victim's coping strategies, whether seeking support, changing their behavior, or even withdrawing, is critical for developing effective support from industry and policymakers.
    \item \textbf{Understanding Vulnerability and Resilience:} As we discuss in \textit{Section 4.2}, while top scam types remain fairly consistent across surveyed countries, scam exposure and victimization rates vary across countries. This suggests that there is a complex interplay of factors contributing to vulnerability and resilience among populations across these countries. Prior research has identified demographic factors (like income, age, education), psychological traits and risky online behavior as key contributors to victimization ~\autocite{Vitak2018-kr, Modic2018-ss, Whitty2020-gw, Norris2019-fi, Modic2013-wn}. We find that while demographic factors like income and internet access are linked to scam victimization, financial losses stemming from victimization is more common in less affluent countries, even when controlling for internet access. This suggests that factors such as cultural attitudes toward risk, financial/digital literacy, and access to information and support could also play a role in resilience, warranting further research.
    \item \textbf{Identifying Barriers to Scam Reporting in countries with high under-reporting:} The under-reporting of scams, particularly in less affluent countries where individuals lack awareness of or access to reporting mechanisms, hinders efforts to understand and address the issue, as highlighted by our results (see \textit{Figure \ref{fig:reported}}). Prior research has shown that shame, stigma, and fear of retaliation or judgment often prevent victims from seeking help or reporting scams, particularly in romance scams~\autocite{Havers2024-ah}. Distrust in authorities and victim-blaming further deter victims from reporting~\autocite{Button2009-wz}. Complexity of reporting processes can also sometimes be a barrier, especially considering victims' post-victimization state of mind~\autocite{Button2013-mz}. It is likely that this problem can be addressed with increased focus on qualitative research in such areas to understand underlying factors leading to high under-reporting in these countries.
\end{enumerate}

\section{Conclusion}

Through a nationally representative, multi-country, scams-focused survey, the present study addresses a substantial gap in our understanding of the global scams landscape. Our findings substantiate recent theories about the role of real-time payments in facilitating scams’ proliferation, and highlight the value of consistent scams reporting across countries to help us understand regional issues in context. By analyzing data from a diverse range of economies and cultures, we contribute important insights for researchers, practitioners, and policymakers in online fraud and scams prevention.

\clearpage
\pretolerance=200
\tolerance=300
\printbibliography
\pretolerance=400
\tolerance=500
\newpage

\section*{Authors}

\textbf{Mo Houtti} is a Student Researcher on the Trust \& Safety Research Team at Google.

\textbf{Abhishek Roy} is a Staff UX Researcher on the Trust \& Safety Research Team at Google.

\textbf{Venkata Narsi Reddy Gangula} is a Senior UX Researcher on the Trust \& Safety Research Team at Google.

\textbf{Ashley Marie Walker} is a Senior UX Researcher on the Trust \& Safety Research Team at Google.

\section*{Disclosures}

This research was funded in it’s entirety by Google. Google is a member of the Global Anti-Scam Alliance, a coalition body of over 100 organizations, including governments, law enforcement, consumer protection, financial authorities and services, social media, internet service providers and cybersecurity professionals.

\end{document}